\documentclass[preprint,amsmath,amssymb,nofootinbib]{revtex4}
\usepackage{graphicx}
\usepackage{dcolumn}
\usepackage{bm}

\begin{document}
\begin{center}
{\large\bf GIANT RESONANCES UNDER EXTREME CONDITIONS}\footnote{Plenary
lecture at the international conference on ``Current problems in nuclear
physics and atomic energy'', May 29 - June 3, 2006, Kiev - Ukraine.}
\end{center}
\begin{center}
{\bf Nguyen Dinh Dang}
\end{center}
\begin{center}
{\it 
    1) Heavy-ion nuclear physics laboratory, Nishina Center for
Accelerator-Based Science, RIKEN, 2-1 Hirosawa, Wako city, 351-0198 
 Saitama, Japan\\
2) Institute for Nuclear Science and Technique, Hanoi, Vietnam
}
\end{center}

    ~\hspace{-0.8cm} {\bf Abstract:} The theoretical description of
    nuclear resonances 
    at zero and finite temperatures is presented.
    The following issues are addressed: 
    
    1) Giant dipole resonances (GDR) in highly excited nuclei, 
    including both low and high regions of temperature.
    The results of calculations are obtained within the phonon-damping 
    model, thermal shape-fluctuation model including thermal 
    pairing, and compared with experimental data.

    2) The electromagnetic cross sections of the 
    double GDRs (DGDR) in $^{136}$Xe and
    $^{208}$Pb. The results obtained in theoretical calculations are 
    compared with the experimental data for the DGDR cross sections  
    in exclusive measurements at near-relativistic energies.

    3) GDR and pygmy dipole resonances (PDR) in neutron-rich nuclei, where 
    the effect of coupling of the GDR to complicated configurations on the PDR is analyzed.
    \\
   \begin{center}
    {\bf 1. Introduction}~
\end{center}
\hspace{5mm}

~Nuclear resonances under extreme conditions mentioned in the present lecture 
include i) giant dipole resonances (GDR) in highly excited nuclei formed in heavy-ion reactions and
inelastic scattering of light particles or nuclei on heavy targets,
ii) double GDRs (DGDR) formed in Coulomb excitations at near
relativistic energies, and iii) pygmy dipole resonances (PDR) in
neutron-rich nuclei.

The recent studies of these resonances have been facing the following
challenges: 

1) The GDR has been observed in highly-excited (hot) nuclei. These nuclei
are formed as compound nuclei at high excitation energies $E^{*}$ in heavy-ion fusion reactions
or in the inelastic scattering of light particles (nuclei) on a heavy 
nucleus. The $\gamma$-decay spectra of these compound nuclei show the existence of the
GDR, whose peak energy depends weakly on the excitation energy $E^{*}$.
The dependence of the GDR on the temperature $T$ has been experimentally
extracted when the angular momentum of the compound nucleus is low, as in the case 
of the light-particle scattering experiments, or when it can be separated out
from the excitation energy $E^{*}$. It has been experimentally found that the GDR's 
full width at the half maximum (FWHM) remains almost constant at
$T\leq$ 1 MeV, but sharply increases with $T$ up to $T\simeq$ 2 - 3 MeV,
and saturates at higher $T$~\cite{Woude}. 

2) The DGDR has been observed in the Coulomb excitations of $^{136}$Xe
and $^{208}$Pb projectiles on $^{208}$Pb target at the bombarding energies of 0.7
GeV/n and 0.64 GeV/n, respectively. The extracted electromagnetic (EM) cross
sections of the DGDR in these experiments have been compared with
theoretical predictions based on the non-interacting phonon (harmonic)
picture. The latter calculates the parameters of the DGDR by folding to independent GDRs. As a
result, the DGDR energy $E_{\rm DGDR}$ is found to be equal to 2$E_{\rm GDR}$
($E_{\rm GDR}$ is the GDR energy), and the DGDR FWHM $\Gamma_{\rm DGDR}$ 
is equal to 2$\Gamma_{\rm GDR}$ ($\Gamma_{\rm GDR}$ is
the GDR FWHM) if Lorentzian photoabsorption cross sections are used in
folding the GDRs, or to $\sqrt{2}\Gamma_{\rm GDR}$, if Gaussians are
folded. The comparison shows that the experimentally extracted energy 
and width of DGDR differ slightly from these values. The most striking
discrepancy is that the experimental EM cross sections are much larger
that than the values predicted by the folding model. The enhancement
is found to be around 178\% for $^{136}$Xe and 133\% for
$^{208}$Pb~\cite{DGDR}.

3) It is well believed that, in neutron-rich nuclei, the oscillation
of the neutron excess against the stable core may cause a low-frequency E1
resonance, which is called the pygmy dipole resonance (PDR). However, in
the case when the neutron excess cannot be well separated from the
stable core, the coupling between them may lead to a strong
fragmentation of the PDR. As a result the pygmy dipole mode may become
non-collective. Theoretical calculations within the relativistic RPA and the 
particle-hole $(ph)\oplus$phonon-coupling
model so far have given contradicting predictions. 
The former shows a clear PDR bump with a peak at around 8 MeV, while
no evidence for collectivity is seen below 10 MeV in the
latter~\cite{PDR}.

It will be shown in the present lecture how the above-mentioned issues
are interpreted within the phonon-damping model (PDM)~\cite{PDM}. The PDM was
originally proposed in 1998 with the primary aim to describe the
damping of the hot GDR. This model was extended later to address the
issue of anharmonicity in the DGDR. By including pairing, the model is
also able to predict the GDR's width at low $T$ as well as the EM cross sections of E1 excitations 
in neutron-rich nuclei.   
\\
\begin{center}
    {\bf 2. Outline of the PDM}
\end{center}    
\hspace{5mm} 

{\it A. Damping of hot GDR}

The quasiparticle representation of the PDM 
Hamiltonian~\cite{PDM1} 
is obtained by adding the superfluid pairing interaction and 
expressing the particle ($p$) and hole ($h$) creation and 
destruction operators, $a_{s}^{\dagger}$ and $a_{s}$ ($s=p, h$), in 
terms of the quasiparticle operators, $\alpha_{s}^{\dagger}$ and
$\alpha_{s}$, using the Bogolyubov's canonical transformation.
As a result, the PDM Hamiltonian 
for the description of E$\lambda$ excitations 
can be written in spherical basis as
\[
H=\sum_{jm}E_{j}\alpha_{jm}^{\dagger}\alpha_{jm}+
\sum_{\lambda\mu i}\omega_{\lambda i}b_{\lambda\mu 
i}^{\dagger}b_{\lambda\mu i}~+
\]
\begin{equation}
\frac{1}{2}\sum_{\lambda\mu i}\frac{(-)^{\lambda-\mu}}{\hat{\lambda}}
\sum_{jj'}f_{jj'}^{(\lambda)}\Biggl\{u_{jj'}^{(+)}
\biggl[A_{jj'}^{\dagger}(\lambda\mu)+A_{jj'}(\lambda\tilde{\mu})\biggr]+
v_{jj'}^{(-)}
\biggl[B_{jj'}^{\dagger}(\lambda\mu)+B_{jj'}(\lambda\tilde{\mu})\biggr]\Biggr\}
\biggl(b_{\lambda\mu i}^{\dagger}+b_{\lambda\tilde{\mu} i}\biggr)~,
\label{H}
\end{equation}
where $\hat{\lambda}=\sqrt{2\lambda+1}$. 
The first term at the right-hand side (rhs) of Hamiltonian (\ref{H})
corresponds to the independent-quasiparticle field. The second term 
stands for the phonon field described by phonon operators, 
$b_{\lambda\mu i}^{\dagger}$ and $b_{\lambda\mu i}$, with 
multipolarity $\lambda$, which generate the harmonic collective
vibrations such as GDR. Phonons are ideal bosons within the PDM, i.e. 
they have no fermion structure. The last term is the coupling between 
quasiparticle and phonon fields, 
which is responsible for the microscopic damping
of collective excitations. 
 
In Eq. (\ref{H}) the following standard notations are used
\begin{equation} 
A_{jj'}^{\dagger}(\lambda\mu)=\sum_{mm'}\langle 
jmj'm'|\lambda\mu\rangle\alpha_{jm}^{\dagger}\alpha_{j'm'}^{\dagger}~,
\hspace{2mm} B_{jj'}^{\dagger}(\lambda\mu)=-\sum_{mm'}(-)^{j'-m'}\langle 
jmj'-m'|\lambda\mu\rangle\alpha_{jm}^{\dagger}\alpha_{j'm'}~,
\label{ABdagger}
\end{equation}
with
$(\lambda\tilde{\mu})\longleftrightarrow(-)^{\lambda-\mu}(\lambda-\mu)$.
Functions $u_{jj'}^{(+)}\equiv u_{j}v_{j'}+v_{j}u_{j'}$ and 
$v_{jj'}^{(-)}
\equiv u_{j}u_{j'}-v_{j}v_{j'}$ are combinations of Bogolyubov's $u$ 
and $v$ coefficients. The quasiparticle energy $E_{j}$ is 
calculated from the single-particle energy $\epsilon_{j}$ as
\begin{equation}
E_{j}=\sqrt{({\epsilon}_{j}'-\epsilon_{\rm F})^{2}+\Delta^{2}}~,
\hspace{1cm} {\epsilon}_{j}'\equiv \epsilon_{j}-Gv_{j}^{2},
\label{epsilonj}
\end{equation}
where the pairing gap $\Delta$ and the Fermi energy $\epsilon_{\rm F}$  
are defined as solutions of the BCS equations. At $T\neq$ 0
the thermal pairing gap $\Delta(T)$ (or $\bar\Delta(T)$) is defined from the
finite-temperature BCS (or modified BCS) equations (See section C
below).

The equation for the propagation of the GDR phonon, which is damped 
due to coupling to the quasiparticle field, is derived making use of 
the double-time Green's function method (introduced by
Bogolyubov and Tyablikov, and developed further by Zubarev). Following the 
standard procedure of deriving the equation for the double-time 
retarded Green's function with respect to the Hamiltonian (\ref{H}), one obtains 
a closed set of equations for the Green's functions for phonon and
quasiparticle propagators. Making the 
Fourier transform into the energy plane $E$, and expressing all the 
Green functions in the set in terms of the one-phonon propagation 
Green function, we obtain the equation for the latter, 
$G_{\lambda i}(E)$, in the form
\begin{equation}
G_{\lambda i}(E)=\frac{1}{2\pi}\frac{1}{E-\omega_{\lambda 
i}-P_{\lambda i}(E)}~,
\label{GE}
\end{equation}
where the explicit form of 
the polarization operator $P_{\lambda i}(E)$
is
\begin{equation}
P_{\lambda i}(E)=\frac{1}{\hat{\lambda}^{2}}\sum_{jj'}[f_{jj'}^{(\lambda)}]^{2}
\biggl[\frac{(u_{jj'}^{(+)})^{2}(1-n_{j}-n_{j'})(\epsilon_{j}+\epsilon_{j'})}
{E^{2}-(\epsilon_{j}+\epsilon_{j'})^{2}}
- \frac{(v_{jj'}^{(-)})^{2}(n_{j}-n_{j'})(\epsilon_{j}-\epsilon_{j'})}
{E^{2}-(\epsilon_{j}-\epsilon_{j'})^{2}}\biggr]~.
\label{PE}
\end{equation}
The polarization operator (\ref{PE}) appears due to $ph$ -- 
phonon coupling in the last term of the rhs of Hamiltonian (\ref{H}).
The phonon damping $\gamma_{\lambda i}(\omega)$ ($\omega$ real) is 
obtained as the imaginary part of the analytic continuation of
the polarization operator $P_{\lambda i}(E)$ into the complex energy 
plane $E=\omega\pm i\varepsilon$. Its final form is
\[
\gamma_{\lambda i}(\omega)=\frac{\pi}{2\hat{\lambda}^{2}}
\sum_{jj'}[f_{jj'}^{(\lambda)}]^{2}\biggl\{
(u_{jj'}^{(+)})^{2}(1-n_{j}-n_{j'})
[\delta(E-E_{j}-E_{j'})-
\delta(E+E_{j}+E_{j'})]-
\]
\begin{equation}
(v_{jj'}^{(-)})^{2}(n_{j}-n_{j'})[\delta(E-E_{j}+E_{j'})
-\delta(E+\epsilon_{j}-\epsilon_{j'})]\biggr\}.
\label{gamma}
\end{equation}
The quasiparticle occupation number $n_{j}$ is 
calculated as 
\begin{equation}
     n_{j}=\frac{1}{\pi}\int_{-\infty}^{\infty}\frac{n_{\rm F}(E)
    \gamma_{j}(E)}{[E-E_{j}-M_{j}(E)]^{2}+\gamma_{j}^{2}(E)}dE~,
    \hspace{6mm} n_{\rm F}(E)=({\rm e}^{E/T}+1)^{-1}~,
     \label{nj}
 \end{equation}
 where $M_{j}(E)$ is 
 the mass operator, and 
 $\gamma_{j}(E)$ is the quasiparticle damping, which is determined as 
 the imaginary part of the complex continuation of $M_{j}(E)$ into the 
 complex energy plane~\cite{PDM1}. These quantities appear 
 due to coupling between quasiparticles and the GDR.  From Eq. (\ref{nj}) it is seen that
 the functional form for the occupation number  $n_{j}$ 
 is not given by the Fermi-Dirac distribution $n_{\rm F}(E_{j})$ 
 for non-interacting 
 quasiparticles. It can be approximately to be so
 if the quasiparticle damping $\gamma_{j}(E)$ is sufficiently small. 
 Equation (\ref{nj}) also implies a zero value 
 for $n_{j}$ in the ground state, i.e. $n_{j}(T=0)=$ 0. 
 In general, it is not the case because of ground-state 
 correlations beyond the quasiparticle RPA (QRPA). They lead to 
 $n_{j}(T=0)\neq$ 0, which should be found by solving 
 self-consistently a set of nonlinear equations within the renormalized 
 QRPA. However, for collective high-lying excitations such as GDR, the value $n_{j}(T=0)$ 
 is negligible. 

The energy $\bar{\omega}$ 
of giant resonance (damped collective phonon) is found 
as the solution of the equation: 
$\bar{\omega}-\omega_{\lambda i}-P_{\lambda i}(\bar{\omega})=0$~.
The width $\Gamma_{\lambda}$ of giant resonance 
is calculated as twice of the damping 
$\gamma_{\lambda}(\omega)$ at $\omega=\bar{\omega}$,
where $\lambda=$ 1 corresponds to the GDR width $\Gamma_{\rm GDR}$. 
The latter is conveniently decomposed into the quantal 
($\Gamma_{\rm Q}$) and thermal ($\Gamma_{\rm T}$) widths 
as
\begin{subequations}
\label{width}
\begin{equation}
\Gamma_{\rm GDR}=\Gamma_{\rm Q}+\Gamma_{\rm T}~,
\label{widthtotal}
\end{equation}
\begin{equation}
\Gamma_{\rm Q}=2\pi F_{1}^{2}\sum_{ph}[u_{ph}^{(+)}]^{2}(1-n_{p}-n_{h})
\delta(E_{\rm GDR}-E_{p}-E_{h})~,
\label{GammaQ}
\end{equation}
\begin{equation}
\Gamma_{\rm T}=2\pi F_{2}^{2}\sum_{s>s'}[v_{ss'}^{(-)}]^{2}(n_{s'}-n_{s})
\delta(E_{\rm GDR}-E_{s}+E_{s'})~,
\label{GammaT}
\end{equation}
\end{subequations}
where $(ss')=(pp')$ and $(hh')$ with $p$ and $h$ denoting the orbital 
angular momenta $j_{p}$ and $j_{h}$ for particles and holes, 
respectively. The quantal and thermal widths come from 
the couplings of quasiparticle pairs 
$[\alpha_{p}^{\dagger}\otimes\alpha^{\dagger}_{h}]_{LM}$ and 
$[\alpha_{s}^{\dagger}\otimes{{\alpha}}_{\widetilde{s'}}]_{LM}$ 
to the GDR, respectively. At zero pairing they correspond to the couplings of 
$ph$ pairs, $[a^{\dagger}_{p}\otimes{{a}}_{\widetilde{h}}]_{LM}$, and 
$pp$ ($hh$) pairs, 
$[a^{\dagger}_{s}\otimes{a}_{\widetilde{s'}}]_{LM}$,
to the GDR, respectively (The tilde $~{\widetilde{}}$~ denotes 
the time-reversal operation). 

The line shape of the GDR is described by the strength function 
$S_{\rm GDR}(\omega)$, which is   
derived from the spectral intensity 
in the standard way using the analytic continuation of the
Green function (\ref{GE}) and
by expanding the polarization operator (\ref{PE}) 
around $\omega=E_{\rm GDR}$. The final form 
of $S_{\rm GDR}(\omega)$ is~\cite{PDM,PDM1}
\begin{equation}
S_{\rm GDR}(\omega)=\frac{1}{\pi}\frac{\gamma_{\rm GDR}(\omega)}
{(\omega-E_{\rm GDR})^{2}+\gamma_{\rm GDR}^{2}(\omega)}~.
\label{S}
\end{equation} 
The photoabsorption cross section $\sigma(E_{\gamma})$ \
is calculated from the strength 
function $S_{\rm GDR}(E_{\gamma})$ as
\begin{equation}
\sigma(E_{\gamma})=
c_{1}S_{\rm GDR}(E_{\gamma})E_{\gamma}~,
\label{sigma}
\end{equation}
where $E_{\gamma}\equiv\omega$ is used to denote the energy of 
$\gamma$-emission. The normalization factor $c_{1}$ is defined 
so that the total integrated photoabsorption cross section 
$\sigma=\int\sigma(E_{\gamma})dE_{\gamma}$ satisfies
the GDR sum rule ${\rm SR}_{\rm GDR}$, 
hence
\begin{equation}
c_{1}={\rm SR}_{\rm GDR}\biggl/\int_{0}^{E_{\rm max}}S_{\rm 
GDR}(E_{\gamma})E_{\gamma}dE_{\gamma}~.
\label{c1}
\end{equation}
In heavy nuclei with $A\geq$ 40, the GDR exhausts the 
Thomas-Reich-Kuhn sum rule (TRK) 
${\rm SR}_{\rm GDR}={\rm TRK}\equiv$ 60 $NZ/A$ (MeV$\cdot$mb) at
the upper integration limit
$E_{\rm max}\simeq$ 30 MeV, and exceeds TRK (${\rm SR}_{\rm GDR} > {\rm 
TRK}$) at $E_{\rm max}>$ 30 MeV due to 
the contribution of exchange forces. In some light nuclei, such as $^{16}$O,
the observed photoabsorption cross section exhausts 
only around 60$\%$ of TRK up to $E_{\rm max}\simeq$ 30 MeV.
\\
\\
{\it B. Thermal pairing}

The standard finite-temperature BCS (FT-BCS) theory ignores
fluctuations of the quasiparticle number. As a result, the BCS breaks 
down at a critical temperature $T_{\rm c}\simeq 0.567\Delta(T=0)$,
which corresponds to the sharp transition from the superfluid phase to
the normal-fluid one. It has been known that, in finite systems such as
nuclei, thermal
fluctuations smooth out this phase transition~\cite{Moretto}.

The modified BCS (MBCS) theory~\cite{MBCS} proposes a microscopic way to include 
quasiparticle-number fluctuations via the secondary
Bogolyubov's transformation
\begin{equation}
\bar{\alpha}_{jm}^{\dagger}=
\sqrt{1-n_{j}}\alpha_{jm}^{\dagger} + 
\sqrt{n_{j}}\alpha_{j\widetilde{m}}~,\hspace{5mm} 
\bar{\alpha}_{j\widetilde{m}}=
\sqrt{n_{j}}\alpha_{j\widetilde{m}} -\sqrt{n_{j}}\alpha_{jm}^{\dagger}~.
\label{alphabar}
\end{equation}
Using Eqs. (\ref{alphabar})  
in combination with the original 
Bogolyubov's transformation, one obtains
the transformation from the particle operators directly to the modified 
quasiparticle operators in the following form
\begin{equation}
a_{jm}^{\dagger}=\bar{u}_{j}\bar{\alpha}_{jm}^{\dagger}+\bar{v}_{j}
\bar{\alpha}_{j\widetilde{m}}~,\hspace{5mm} 
a_{j\widetilde{m}}=\bar{u}_{j}\bar{\alpha}_{j\widetilde{m}}
-\bar{v}_{j}\bar{\alpha}_{j{m}}^{\dagger}~,
\label{ajm}
\end{equation}
where the coefficients $\bar{u}_{j}$ and $\bar{v}_{j}$ are related
to the conventional Bogolyubov's coefficients $u_{j}$ and $v_{j}$ as
\begin{equation}
\bar{u}_{j}=u_{j}\sqrt{1-n_{j}}+v_{j}\sqrt{n_{j}}~,
\hspace{5mm} 
\bar{v}_{j}=v_{j}\sqrt{1-n_{j}}-u_{j}\sqrt{n_{j}}~.
\label{ubarvbar}
\end{equation}    
The transformation of the pairing Hamiltonian
(\ref{H}) into the modified quasiparticles 
$\bar{\alpha}_{jm}^{\dagger}$ and $\bar{\alpha}_{jm}$ has the form 
identical to that obtained within the conventional quasiparticle
representation with ($\bar{u}_{j}$, $\bar{v}_{j}$) 
replacing ($u_{j}$, $v_{j}$) and ($\bar{\alpha}_{jm}^{\dagger}$, 
$\bar{\alpha}_{jm}$) replacing (${\alpha}_{jm}^{\dagger}$, 
${\alpha}_{jm}$), respectively. 
The MBCS equations, therefore, has exactly the same form as 
that of the standard BCS equations, 
where the coefficients $u_{j}$ and $v_{j}$ are replaced with
$\bar{u}_{j}$ and $\bar{v}_{j}$, i.e.
\begin{equation}
{\bar{\Delta}}=G\sum_{j}\Omega_{j}\bar{u}_{j}\bar{v}_{j}=G\sum_{j}\Omega_{j}[(1-2n_{j})u_{j}v_{j}-
\sqrt{n_{j}(1-n_{j})}(u_{j}^{2}-v_{j}^{2})]~,
\label{MBCSgap}
\end{equation}
\begin{equation}
N=2\sum_{j}\Omega_{j}\bar{v}_{j}^{2}=2\sum_{j}\Omega_{j}[(1-2n_{j})v_{j}^{2}+n_{j}
-2\sqrt{n_{j}(1-n_{j})}u_{j}v_{j}]~,
\label{MBCSnumber}
\end{equation}
The last terms at the rhs of these MBCS equations contain the
quasiparticle-number fluctuations $\sqrt{n_{j}(1-n_{j})}$ on $j-$th
orbitals, which are not included in the standard FT-BCS theory. 
\\
\\
{\it C. EM cross sections of GDR and DGDR}

The EM cross section
$\sigma_{\rm EM}$
is calculated
from the corresponding photoabsorption cross section 
$\sigma(E_{\gamma})$ and the photon spectral function $N(E_{\gamma})$ as
\begin{equation}
\sigma_{\rm EM}=
\int N(E_{\gamma})\sigma(E_{\gamma})dE_{\gamma},\hspace{5mm} 
N(E_{\gamma})=2\pi\int_{b_{\rm min}}^{\infty}{\rm 
e}^{-m(b)}N(E_{\gamma},b)bdb.
\label{sigmaEM}
\end{equation}
The expression for the spectrum $N(E_{\gamma},b)$ of virtual photons 
from a stationary 
target as seen by a projectile moving with a velocity 
$\beta=v/c$ at the impact parameter $b$ is also given in~\cite{Pshe}.
The average number of photons absorbed by the projectile 
is calculated as $m(b)=\int_{E_{\rm 
min}}^{\infty}N(E_{\gamma},b)\sigma(E_{\gamma})dE_{\gamma}$~.

The DGDR strength function is calculated within the PDM 
as
\begin{equation}
S_{\rm DGDR}^{\rm PDM}(E)=\frac{2}{\pi}
\frac{\gamma_{\rm DGDR}(E)}{(E-E_{\rm DGDR})^{2}+
[\gamma_{\rm DGDR}(E)]^{2}},
\label{SDGDR}
\end{equation}
where the DGDR energy $E_{\rm DGDR}$ and damping
$\gamma_{\rm DGDR}(E)$ 
are calculated microscopically within the PDM (See details in Ref. \cite{DTA}).  
The DGDR cross section
$\sigma_{\rm DGDR}(E)$ is calculated as
\begin{equation}
\sigma_{\rm DGDR}(E)=c^{(2)}S_{\rm DGDR}^{\rm PDM}(E)E~.
\label{sigmaDGDR}
\end{equation}
The DGDR 
strength factor $c^{(2)}$ in (\ref{sigmaDGDR})
as follows. Using (\ref{sigmaEM}) and the harmonic limit 
$S_{\rm DGDR (har)}^{\rm PDM}(E)$ of the DGDR 
strength function (\ref{SDGDR}), which is obtained
by folding two GDR strength functions (\ref{S}) (pairing not included), we write the formal expression 
of the harmonic limit $\sigma_{\rm C}^{(2)}(\rm har)$ 
of the EM 
cross section (\ref{sigmaEM}) for DGDR as
\begin{equation}
\sigma_{\rm C}^{(2)}({\rm har})=
\int\frac{d\sigma_{\rm C}^{(2)}}{dE}({\rm har})dE=
c^{(2)}\int N_{\rm har}(E)
S_{\rm DGDR(har)}^{\rm PDM}(E)EdE,
\label{sigma2har}
\end{equation}
where $N_{\rm har}(E)$ is calculated using 
the harmonic limit $\sigma_{\rm DGDR(har)}^{PDM}(E)$ of 
(\ref{sigmaDGDR}) in 
(\ref{sigmaEM}) and $m(b)$. We require
this cross section (\ref{sigma2har}) to be equal
to the one calculated by folding two GDR 
cross sections, namely
\[
\sigma_{\rm C(f)}^{(2)}\equiv
\int\frac{d\sigma_{\rm C(f)}^{(2)}}{dE}dE=
\frac{1}{2}\int dEdE_{1}dE_{2}
N(E_{1},E_{2})\sigma_{\rm GDR}(E_{1})
\sigma_{\rm GDR}(E_{2})\delta(E-E_{1}-E_{2})=
\]
\begin{equation}
\frac{[c^{(1)}]^{2}}{\pi}\int dEdE_{1}dE_{2}
N(E_{1},E_{2})S_{\rm GDR}^{\rm PDM}(E_{1})
S_{\rm GDR}^{\rm PDM}(E_{2})E_{1}E_{2}
{\varepsilon}/[(E-E_{1}-E_{2})^{2}+\varepsilon^{2}],
\label{folding}
\end{equation}
where the representation $\delta(x)=
[(x-i\varepsilon)^{-1}-(x+i\varepsilon)^{-1}]/(2\pi i)$
and the expression for $N(E_{1},E_{2})$ given in
\cite{Pshe} are used. Equalizing the right-hand sides
of (\ref{sigma2har}) and (\ref{folding}), 
we define $c^{(2)}$. Knowing $c^{(2)}$, 
we can calculate the
EM cross section $\sigma_{\rm C}^{(2)}$
of the DGDR from (\ref{sigma2har}) using 
$S_{\rm DGDR}^{\rm PDM}(E)$ (\ref{SDGDR})
instead of its harmonic limit.
\\
\begin{center}
    {\bf 3. Numerical results}
    \end{center}
    \hspace{5mm}
 
{\it A. Assumptions and parameters of PDM}

The PDM is based on the following assumptions:

a1) The matrix elements for the coupling of GDR to non-collective $ph$ 
configurations, which causes the quantal width $\Gamma_{\rm Q}$ 
(\ref{GammaQ}), are all equal to $F_{1}$. 
Those for the coupling of GDR to $pp$ ($hh$), which causes the thermal 
width $\Gamma_{\rm T}$ (\ref{GammaT}), 
are all equal to $F_{2}$.

a2) It is well established that 
the microscopic mechanism of the quantal (spreading) width 
$\Gamma_{\rm Q}$ (\ref{GammaQ}) comes from 
quantal coupling of $ph$ configurations to more complicated ones, 
such as $2p2h$ ones. The calculations performed 
in Refs. \cite{NFT} 
within two independent microscopic models, where such couplings to 
$2p2h$ configurations were explicitly included, have shown that $\Gamma_{\rm Q}$ 
depends weakly on $T$. Therefore, 
in order to avoid complicate numerical calculations, which are not 
essential for the increase of $\Gamma_{\rm GDR}$ 
at $T\neq$ 0, such microscopic mechanism is not 
included within PDM, assuming
that $\Gamma_{\rm Q}$ at $T=$ 0 is known. The model parameters
are then chosen so that the calculated $\Gamma_{\rm Q}$ and $E_{\rm 
GDR}$ reproduce the corresponding experimental values 
at $T=$ 0. 

Within assumptions (a1) and (a2) the model has only three 
$T$-independent parameters, which are the unperturbed phonon energy 
$\omega_{q}$, $F_{1}$, and  $F_{2}$. 
The parameters $\omega_{q}$ and $F_{1}$ are 
chosen so that after the $ph$-GDR coupling is switched on,
the calculated GDR energy $E_{\rm GDR}$ and width $\Gamma_{\rm GDR}$ 
reproduce the corresponding experimental values for GDR on 
the ground-state. At $T\neq$ 0, the coupling to $pp$ and $hh$ 
configurations is activated.
The $F_{2}$ parameter is then fixed at $T=$ 0 
so that the GDR energy $E_{\rm GDR}$ 
does not change appreciably with $T$. 

{\it B. Temperature dependence of GDR width}
\begin{figure}                                                             
\includegraphics[width=13cm]{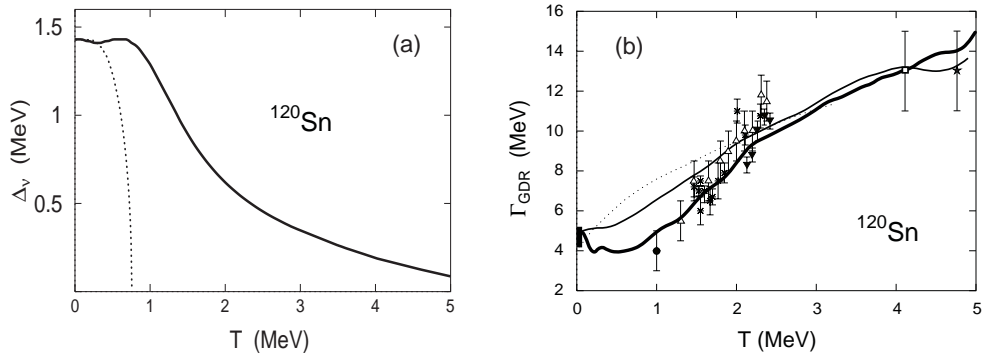}
\caption{\label{widthSn}(a): Neutron pairing gap 
for $^{120}$Sn as a function of $T$.
Solid and dotted lines show the MBCS and BCS 
gaps, respectively. (b): GDR width $\Gamma_{\rm GDR}$ as a function of 
$T$ for $^{120}$Sn. The thin and thick solid lines show the 
PDM results obtained neglecting pairing and  
including the renormalized gap $\widetilde{\Delta}=[1+1/\delta N^2]\bar{\Delta}$, respectively. 
The gap $\widetilde\Delta$ includes the correction 
$\delta
N^{2}=\bar{\Delta}(0)^{2}\sum_{j}(j+1/2)/[(\epsilon_{j}-\bar{\epsilon}_{\rm
F})^{2}+\bar{\Delta}(0)^{2}]$ due to an
approximate number projection.
The prediction by the TFM is shown as the 
dotted line~\cite{TFM}.}
\end{figure}

Shown in Fig. \ref{widthSn} (a) is the $T$ dependence of the neutron 
pairing gap $\bar{\Delta}_{\nu}$ for $^{120}$Sn, which is 
obtained from the MBCS equation (\ref{MBCSgap}) using the 
single-particle energies 
determined within the Woods-Saxon potential at $T=$ 0. 
The pairing parameter $G_{\nu}$ is chosen to be equal to 0.13 MeV, which 
yields $\bar{\Delta}(T=0)\equiv\bar{\Delta}(0)\simeq$ 1.4 MeV. 
Contrary to the BCS gap (dotted line), 
which collapses at $T_{\rm c}\simeq$ 
0.79 MeV, the 
gap $\bar{\Delta}$ (solid line) does not vanish, but
decreases monotonously with increasing $T$ at $T\agt$ 1 MeV 
resulting in a long tail up to $T\simeq$ 5 MeV.
This behavior is caused by the thermal fluctuation 
of quasiparticle number in the MBCS equations (\ref{MBCSgap}) and
(\ref{MBCSnumber}).

The GDR widths as a function of $T$ for $^{120}$Sn obtained within the
PDM are compared in Fig. \ref{widthSn} (b) with the experimental data and the prediction by the
thermal fluctuation model (TFM)~\cite{TFM}. The TFM interprets the
broadening of the GDR's width via an adiabatic coupling of GDR 
to quadrupole deformations induced by thermal 
fluctuations. Even when thermal pairing is
neglected the PDM prediction, (the thin solid line) is already better than 
that given by the TFM, including the region of high $T$ where the
width's saturation is reported. The increase of the total width with
$T$ is driven by the increase of the thermal width $\Gamma_{\rm T}$
(\ref{GammaT}), which is caused by coupling to $pp$ and $hh$
configurations, since the quantal width $\Gamma_{\rm Q}$ (\ref{GammaQ}) 
is found to decrease slightly with increasing $T$. 
The inclusion of thermal pairing,
which yields a sharper Fermi surface, compensates
the smoothing of the Fermi surface with increasing $T$. This leads to 
a much weaker $T$-dependence of the GDR's width at low $T$. 
As a result, the values of the width predicted by the
PDM in this region significantly drop (the thick solid line),
recovering the data point at $T=$ 1 MeV. 

\begin{figure} 
\includegraphics[width=11cm]{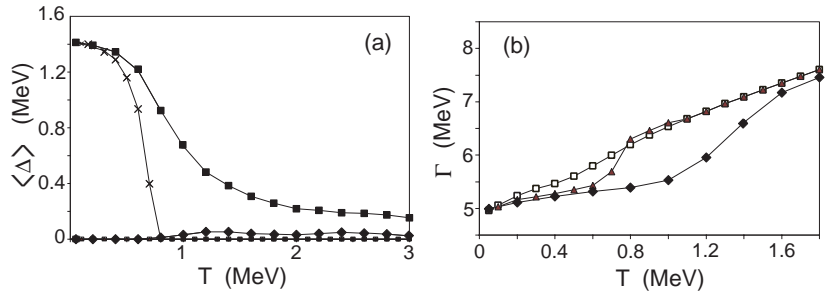}
\caption{\label{Sn120}
(a): Pairing gaps for $^{120}$Sn averaged over thermal shape 
fluctuations versus $T$. 
Lines with triangles and crosses are the usual BCS proton and neutron 
pairing gaps, respectively, while those
with diamonds and squares denote the corresponding pairing gaps, which 
also include thermal fluctuations of pairing fields.
(b): GDR widths for $^{120}$Sn versus $T$. Open squares, 
triangles, and diamonds denote the widths obtained without pairing, 
including BCS pairing, and thermally fluctuating pairing field from (a), 
respectively.}
\end{figure}
The results discussed above have 
also been confirmed by our recent calculations within a macroscopic approach, which takes 
pairing fluctuations into account along with the thermal shape 
fluctuations~\cite{Aru}. Here the free energies 
are calculated using the Nilsson-Strutinsky method at $T\neq$
0, including thermal pairing correlations.  
The GDR is coupled  to the nuclear shapes through a simple 
anisotropic harmonic oscillator model with a separable dipole-dipole 
interaction.  The observables are averaged over the shape parameters and pairing gap. 
Our study reveals that the observed quenching of GDR width at low $T$ in
$^{120}$Sn and $^{148}$Au can be understood in terms of simple 
shape effects caused by pairing correlations.  Fluctuations in
pairing field lead to a slowly vanishing pairing gap [Fig. \ref{Sn120}
(a)], which influences the
structural properties even at moderate $T$ ($\sim$1 MeV).  We
found that the low-$T$ structure and hence the GDR width are 
quite sensitive to the change of the pairing field [Fig. \ref{Sn120}
(b)].  

{\it C. EM cross section of DGDR}

The calculations employ the  
single-particle energies for $^{136}$Xe and $^{208}$Pb 
obtained within the Hartree-Fock (HF)
method using the SGII interaction. 
The unperturbed phonon energy $\omega_{q}$ and 
the degenerate $ph$ matrix element 
$F_{ph}$ are chosen so that the GDR energy $E_{\rm GDR}$ 
and FWHM $\Gamma_{\rm GDR}$, obtained within the PDM, 
reproduce the experimentally extracted values.

\begin{figure}
    \includegraphics[width=10cm]{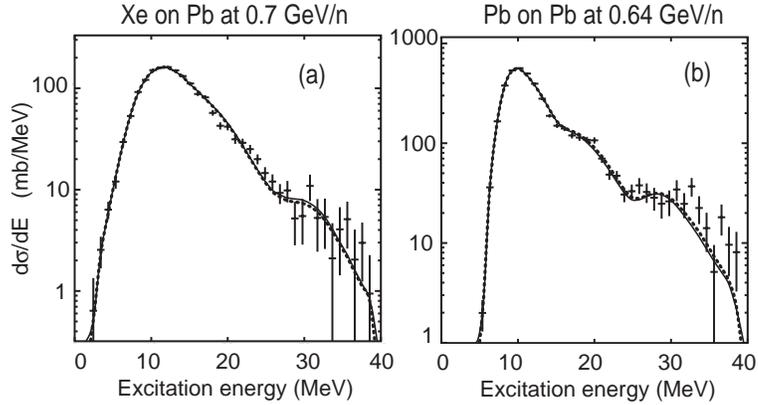}
    \caption{\label{DGDR} Differential EM cross section for $^{136}$Xe 
    (a) and $^{208}$Pb (b). Results obtained using EM cross section 
    of DGDR within PDM (solid lines) and the best fit with $\chi^{2}$ to the data points 
(dotted lines) are based on a normalization of GDR (the 
bump at $\sim$ 10 MeV) which exhausts 90$\%$ of TRK.}
\end{figure}

The peak energies $E_{\rm i}$, FWHM $\Gamma_{\rm i}$, 
and EM cross section 
$\sigma_{\rm C}^{\rm i}$ for GDR (i$=$GDR) 
and DGDR (i$=$DGDR) predicted by the PDM are shown in Table \ref{tab1}  
in comparison
with the experimental data  
for $^{136}$Xe and $^{208}$Pb~\cite{DGDR}. It is seen that all the calculated values are in reasonable agreement
with the corresponding experimental values.
\begin{table}
\caption{The energies $E_{\rm i}$, 
FWHM $\Gamma_{\rm i}$, and EM cross section 
$\sigma_{\rm C}^{\rm i}$ for GDR (i$=$GDR) (a) 
and DGDR (i$=$DGDR) (b), 
calculated within PDM (Theory) 
in comparison
with the experimental data (Exper.) 
for $^{136}$Xe and $^{208}$Pb\label{tab1}}
\begin{tabular}{c|rl|rl|rl}
{a}&\multicolumn{2}{c|}{$E_{\rm GDR}$ (MeV)}
&\multicolumn{2}{c|}{$\Gamma_{\rm GDR}$ (MeV)}
&\multicolumn{2}{c}{$\sigma_{\rm C}^{\rm GDR}$ (mb)~~}\\
\hline
Nucleus
&Theory&~~Exper.~~~~
&Theory&~~Exper.~~~~
&Theory&~~Exper.~~~~\\
\hline
$^{136}$Xe& 15.6
&~~15.2
& 4.96
&~~4.8
& 1676.28
&~~ 1420(42)$\pm 100$\\ 
$^{208}$Pb
& 13.5 
&~~13.4
& 4.04
&~~4.0
& 3039.67 
& ~~3280$\pm 50$\\ 
\hline\hline
{b}&\multicolumn{2}{c|}{$E_{\rm DGDR}$ (MeV)}
&\multicolumn{2}{c|}{$\Gamma_{\rm DGDR}$ (MeV)}
&\multicolumn{2}{c}{$\sigma_{\rm C}^{\rm DGDR}$ (mb)~~}\\
\hline
Nucleus&Theory&~~Exper.~~~~&Theory
&~~Experiment~~~~&Theory&~~Exper.~~~~\\
\hline
$^{136}$Xe
& 29.2
& ~~28.3 $\pm 0.7$
& 7.0
& ~~6.3 $\pm 1.6$
& 159.33
& ~~164(85)$\pm 35$\\ 
$^{208}$Pb
& 26.6
& ~~26.6 $\pm 0.8$
& 6.3
& ~~6.3 $\pm 1.3$
& 420.92
& ~~380 $\pm 40$ 
\end{tabular}
\end{table}
K. Boretzky of the LAN collaboration has folded 
the PDM strength functions for GDR and DGDR 
with the detector response and plotted the obtained results against the experimental fits in Fig.
\ref{DGDR}, which shows a remarkable agreement between the PDM
predictions and the experimental data. 

{\it D. E1-excitations in neutron-rich nuclei}

The calculations of photoabsorption and EM cross sections have been 
carried out for oxygen isotopes with $A =$ 16, 18, 20, 22, and 24,
and for calcium isotopes with $A =$ 40, 42, 44, 46, 48, 50, 52, and 60.
The calculations employ the spherical-basis 
single-particle energies $E_{j}$ obtained
within the HF method using the SGII interaction. 
The two PDM parameters $\omega_{\lambda}$ 
($\lambda=$ 1) and $F_{1}=f_{jj'}^{(1)}$ for all $ph$ indices $(j=p,j'=h)$ 
are chosen in such a way that the values of GDR energy $E_{\rm 
GDR}$
and width $\Gamma_{\rm GDR}$ for $^{16}$O and $^{40,48}$Ca reproduce their 
corresponding experimental values $E_{\rm GDR}^{\rm exp}$ 
and $\Gamma_{\rm GDR}^{\rm exp}$. These chosen values of PDM parameters are then fixed 
in calculations for the neighbor neutron-rich isotopes ($N \geq Z$). 
The neutron pairing gap $\Delta_{\rm n}$ is adjusted around the general 
trend $12/\sqrt{A}$ of the observed pairing gaps in stable nuclei 
to keep the GDR energy stable against varying the neutron number $N$.

\begin{figure}
    \includegraphics[width=10cm]{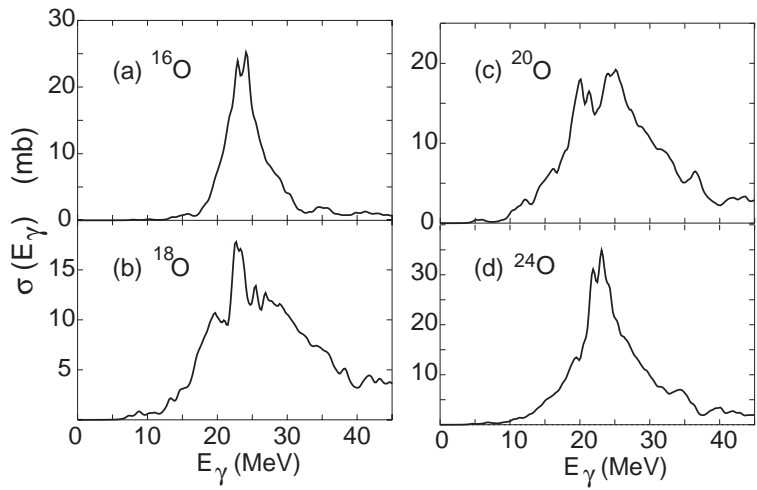}
    \caption{\label{photoO} Photoabsorption cross sections obtained 
within the PDM for some oxygen isotopes.}
\end{figure}
Shown in Fig. \ref{photoO}  
are the photoabsorption cross sections obtained within the PDM for $^{16,18,20,24}$O. 
It is seen 
that the GDR becomes broader with increasing
the neutron number $N$. Its width is particularly large for $^{18, 20}$O,
for which the values of neutron pairing gap $\Delta_{\rm n}$ 
are largest. This increase of GDR spreading enhances both of its
low- and high-energy tails. In the region $E_{\rm\gamma}\leq$ 
15 MeV, some weak structure of PDR is visible for
$^{18,20}$O, and also $^{22}$O (not shown). 
In the rest of isotopes under study, except for an 
extension of the GDR tail toward lower-energy, there is no visible structure
of the PDR.

\begin{figure}
    \includegraphics[width=12cm]{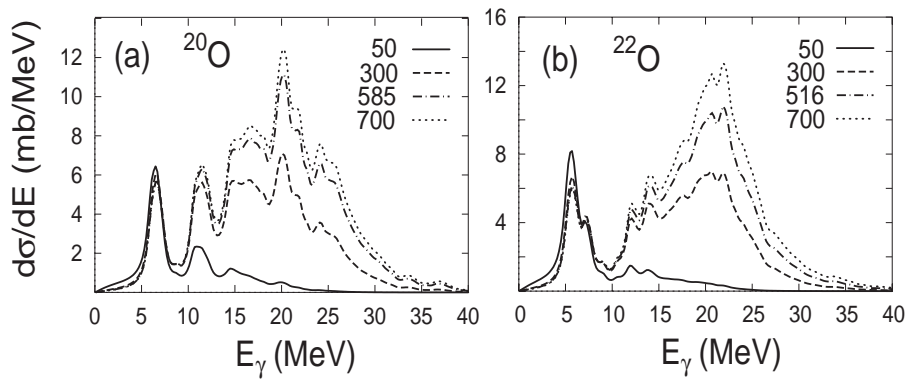}
    \caption{Electromagnetic cross sections of GDR 
within PDM for $^{20,22}$O on $^{208}$Pb target. 
Different lines display results
obtained at different projectile energies as indicated in the panels.
\label{EMO}}
\end{figure}
Since the photon spectral function $N(E_{\rm\gamma})$ 
in the EM differential cross 
section $d\sigma_{\rm EM}/dE_{\rm\gamma}$ (Eq. (\ref{sigmaEM})) is an 
exponentially decreasing function with increasing $E_{\rm\gamma}$, it 
enhances the low-energy part of the GDR. Therefore, 
the  EM differential cross 
section $d\sigma_{\rm EM}/dE_{\rm\gamma}$ of GDR can be used as a 
magnifying glass for the structure
of PDR.  These cross sections obtained within the PDM are 
shown in Fig. \ref{EMO} for $^{20,22}$O. The calculations were carried out for $^{208}$Pb target 
at various projectile energies as shown in these figures.
The PDR shows up in the EM cross sections as a well
isolated peak located at around 6 MeV.  Varying the 
projectile energy affects mostly the GDR region (above 10 MeV) of the
EM cross section, but not
the PDR. 
\\
\begin{center}
    {\bf 4. Conclusions}
    \end{center}
    \hspace{5mm}
 
    A review is given on the current status of the study of nuclear resonances.
    It is demonstrated that  the PDM is a simple but yet microscopic model, which is able
     to describe a variety of resonances under extreme conditions, namely:
     1)  the GDR at $T\neq$ 0, including 
    the constant width at T below 1 MeV, the width increase at low T, the width saturation at high T; 
    2) the EM cross sections for DGDR obtained at near-relativistic energy 
    in $^{136}$Xe and $^{208}$Pb; 3) the possibility of extracting PDR in the EM cross sections of neutron-rich
    isotopes at low-energy beams.
    
    Regarding the GDR at $T\neq$ 0, it is shown that the mechanism of the width's 
    increase at 1 $\leq T\leq$ 3 MeV and its saturation at $T>$ 3 MeV comes from the coupling of the
    GDR to non-collective $pp$ and $hh$ configurations at $T\neq$ 0.
    Meanwhile this effect is nearly cancelled by the monotonous
    decrease of the thermal pairing with increasing $T$ at $T<$ 1 MeV.
    As the result, the GDR width in this low-$T$ region remains nearly 
    temperature independent. This effect is confirmed by calculations 
    in both of the PDM and a macroscopic approach, which takes into account
    thermal fluctuations of nuclear shapes and pairing field.
    
    Concerning the DGDR, the PDM has succeeded to include the effect of
    anharmonicity between two coupled GDRs. As a result, for the first
    time, the
    experimental values of EM cross sections are nicely reproduced by 
    theoretical calculations within the PDM for both $^{136}$Xe and $^{208}$Pb.
    
    Finally, the results of calculations within the PDM have
    demonstrated how the pygmy dipole mode can be depleted due to its coupling to the GDR. 
    This can be the case when the neutron excess cannot be well
    separated from the stable core. This effect leads
    to the disappearance of collectivity of the GDR. As a result the
    photoabsorption cross sections for neutron-rich isotopes
    have a tail extended toward low-energy region instead of a
    well-pronounced PDR peak. Nonetheless, since the photon spectral function $N(E_{\rm\gamma})$ 
in the EM differential cross is an 
exponentially decreasing function with increasing $E_{\rm\gamma}$, it 
enhances the low-energy part of the E1-strength distribution, which is insensitive to the
variation of beam energy. This feature suggests that 
a clean PDR peak (without admixture with the GDR) 
can be seen in the EM cross section (of neutron-rich oxygen and calcium isotopes, e.g.) 
using low-energy, but high-intensity beams at $\sim$ 50 Mev/n.

\end{document}